# Single crystals of the layered dichalcogenides $MoS_2$ and $WS_2$ grown by liquid phase transport

*F. Alex Cevallos[1], Shu Guo[1], Hoseok Heo[2], Giovanni Scuri[2], You Zhou[2], Jiho Sung[2], Takashi Taniguchi[3], Kenji Watanabe[3], Philip Kim[4], Hongkun Park[2], and Robert J. Cava[1]*

[1] Department of Chemistry, Princeton University, Princeton NJ 08542

[2] Department of Chemistry and Chemical Biology, Harvard University, Cambridge MA 02138

[3] National Institute for Materials Science, 1-1 Namiki, Tsukuba 305-0044, Japan

[4] Department of Physics, Harvard University, Cambridge MA 02138

ABSTRACT: The growth of single crystals of $MoS_2$ and $WS_2$ by materials transport through a liquid salt flux made from a low melting mixture of NaCl and CsCl is described. The crystals are characterized by single crystal X-ray diffraction, which reveals that they are 2H-$MoS_2$ with a small percentage (about 3%) of 3R intergrowths. The 2H-$WS_2$ crystals display less than 1% of 3R intergrowths. Photoluminescence spectra of exfoliated monolayers suggest that $MoS_2$ grown by this method has superior crystallinity in comparison to commercially-available $MoS_2$ crystals.

**Introduction**

Transition metal dichalcogenides (TMDCs) are compounds with a composition $MX_2$ where $M$ is a transition metal such as Tungsten or Niobium, and $X$ is a chalcogen such as Sulfur, Selenium or Tellurium. TMDCs have been a subject of active study for several decades due to their wide variety of observed properties [1], including superconductivity [2-4], charge density waves [5, 6], and photoluminescence [7, 8]. In recent years, layered TMDCs such as $MoS_2$ have received intense focus due to the ease with which they can be exfoliated into monolayers, and the interesting properties that result [8, 9].

One issue that has arisen in the study of $MoS_2$ is the difficulty of growing large, good quality single crystals. For example, while $MoS_2$ monolayers can be reliably grown using chemical vapor deposition [10, 11], bulk growth has for the most part been achieved solely by the vapor transport method [12], which often can produce a significant number of defects and atom deficiencies, resulting in significantly different properties under slightly altered growth conditions [13, 14]. In the case of $MoS_2$ in particular, a significant amount of research work has been performed on the naturally-occurring mineral form of the compound, which can result in dramatically varying properties depending on a number of uncontrollable factors [15]. Bearing this in mind, as well as that natural single crystals of $MoS_2$ are an exhaustible and non-renewable resource, the search for additional crystal growth methods for this compound and others is an active field of research.

Recently, Zhang et al. [16] have published an alternative approach to the growth of bulk crystals of $MoS_2$, using a metal "solution transport" method, a horizontal flux growth analogous to the more common vapor transport method [17] that has been previously successfully applied to chalcogenide and pnictide crystal growth [18-20]. The resulting crystals, grown in Tin flux, are high-quality and have many desirable properties, suggesting that solution transport may be a successful general approach to bulk growth of layered TMDCs. Here we employ a low-melting mixture of NaCl and CsCl as a salt flux for the growth of layered $MoS_2$, and extend the application of this technique to the closely-related compound $WS_2$.

**Experimental**

Polycrystalline $MoS_2$ was synthesized by first mixing a stoichiometric amount of Molybdenum powder (99.9%, Alfa Aesar) with 5% excess of Sulfur (99.5%, Johnson Matthey). The mixture was loaded into a sealed, evacuated quartz tube and heated to 950°C at a rate of 60°C/hour. The tube was heated for 10 days, then cooled to room temperature over 24 hours. The resulting material took the form of large, loosely-packed clumps of crystalline $MoS_2$ with a distinct "glitter" appearance, although no single crystals could be isolated of sufficient quality for structural characterization. Polycrystalline $WS_2$ was similarly synthesized, with the substitution of Tungsten powder (99.9%, Alfa Aesar) and a reaction temperature of 1000°C, which resulted in a dull grey powder.

For single crystal growth, a 100 mg pellet of polycrystalline $MoS_2$ or $WS_2$ was placed in a quartz tube of outer diameter 16 mm, wall thickness 0.8 mm, and approximately 7 cm in length (after sealing). A 10 gram mixture of 38 mol% NaCl (99.0%, Alfa Aesar) and 62 mol% CsCl (99%, Alfa Aesar) was used to fill the remaining space in the tube. This salt mixture forms a low-melting eutectic and is reliably liquid at temperatures above 600°C. The tube was sealed under vacuum and placed in a small box furnace in a horizontal configuration, with the end containing the pellet pointed towards the back of the furnace and the growth end pointing towards the furnace door; thus using the natural temperature gradient in the furnace. The tube was then heated to 1000°C ($MoS_2$) or 1100°C ($WS_2$) at 90°C/hour. The molten flux covered the entire bottom of the tube, allowing for "liquid transport" analogous to the more common vapor transport crystal growth method. A diagram of this reaction orientation can be viewed in **Figure 1.** After one week, the tube was cooled to 650°C at a rate of 2°C/hour. During cooling, the molten salt flux remained transparent and crystals could be observed growing on the end of the tube closest to the furnace door. The tube was then removed from the furnace and quenched in a vertical orientation, in order to separate the salt flux from the resulting crystals. The crystals were then washed with distilled water in order to remove the remaining salt. The final products took the form of thin, leafy flakes of silver-colored material with a significant variation in size, although a small number of samples of both materials with a diameter of 5 mm or greater were present.

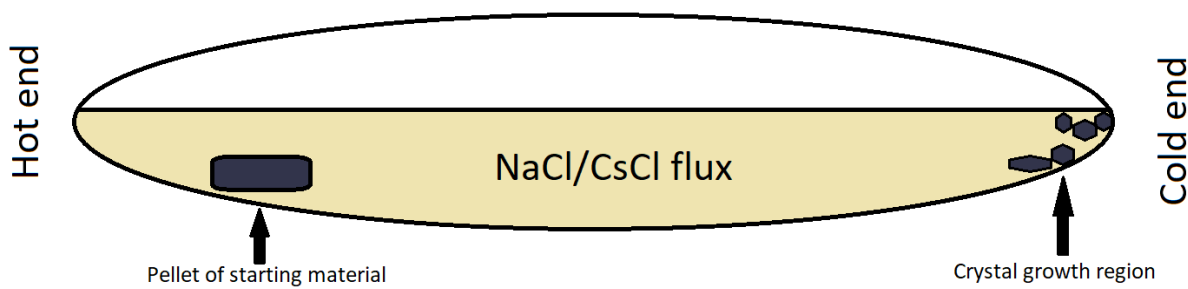

**Figure 1.** Cartoon schematic of the orientation used in solution-transport crystal growth.

Room temperature powder X-ray diffraction (PXRD) measurements were taken using Bruker D8 Advance Eco diffractometer using Cu K$\alpha$ radiation ($\lambda$ = 1.5418 Å) and a LynxEye-XE detector. Single crystal X-ray diffraction (SXRD) data were collected at 300(1) K with a Kappa Apex2 CCD diffractometer (Bruker) using graphite-monochromated Mo-K$\alpha$ radiation ($\lambda$ = 0.71073 Å). The raw data were corrected for background, polarization, and the Lorentz factor and multi-scan absorption corrections were applied. Finally, the structures were analyzed by the Intrinsic Phasing method provided by the ShelXT structure solution program [21] and refined using the ShelXL least-squares refinement package with the Olex2 program. [22] The ADDSYM algorithm in program PLATON was used to double check for possible higher symmetry. [23] Scanning electron microscopy (SEM) was performed using a Quanta 200 FEG environmental (FEI) scanning electron microscope, with an accelerating voltage of 20 KeV. Energy dispersive X-ray spectroscopy (EDX) was performed using an Oxford Instruments X-Max 80 mm$^2$ SDD spectrometer, and data was collected and analyzed through the INCA suite of programs. Atomic force microscopy (AFM) was performed using a Bruker Dimension Icon with ScanAsyst.

For photoluminescence (PL) measurements, thin hexagonal boron nitride (hBN) layers (50 – 80-nm thick) and monolayer $MoS_2$ were mechanically exfoliated from bulk crystals onto $SiO_2$ (285 nm)/Si substrates, and were identified based on their optical contrast. The thickness of the selected hBN flakes was determined by atomic force microscopy measurements. Prior to exfoliation, the $SiO_2$/Si substrates were first ultrasonically cleaned in acetone and 2-propanol for 5 minutes. Following the solution-cleaning step, the substrates were subjected to further cleaning in oxygen plasma (100 W, 300 mTorr) for 10 minutes. Afterwards, the flakes were exfoliated by the scotch-tape method. The hBN/$MoS_2$/hBN heterostructures were then stacked and transferred using a dry transfer method [24] onto cleaned $SiO_2$/Si substrates. PL measurements were carried out in a home-built confocal microscope using an objective with a numerical aperture of 0.75 under 532 nm (2.33 eV) laser excitation in a cryostat from Montana Instruments.

**Results and Discussion**

Single-crystal X-ray diffraction patterns were measured for several planar single-crystals of both $MoS_2$ and $WS_2$. The resulting crystal structures can be seen in **Figure 2**. Crystallographic data and atomic positions for our $MoS_2$ and $WS_2$ can be found in the supporting information. Crystallographic analyses from the positions of sharp Bragg reflections reveal that both compounds crystalize in the hexagonal space group $P6_3/mmc$ (No. 194). Within the unit cells of both $MoS_2$ and $WS_2$, the refinements of the average structures (i.e. what the single crystal diffraction patterns determine) show that there are two unique metal atom sites (Wyckoff sites $2b$ and $2c$). The $2b$ site in both compounds is only slightly occupied (3.14% in $MoS_2$ and 0.79% in $WS_2$), which is an indication of the presence of stacking faults. Similarly, the unit cells of $MoS_2$ and $WS_2$ exhibit two unique S atoms sites (Wyckoff sites $4e$ and $4f$). The $4e$ site in $MoS_2$ is 3.14% occupied, corresponding well to the fraction of Mo atoms in the $2b$ site. As these two values were refined independently and not fixed to each other, their agreement is significant. For $WS_2$, where the scattering factor for S is small compared to that for W, and the fraction of misplaced W's is very small (0.79%), the identification of the associated S position is not as clear as in the $MoS_2$ case, and so its fractional site occupancy is fixed to be equal to that of the W that is out of place due to the stacking faults.

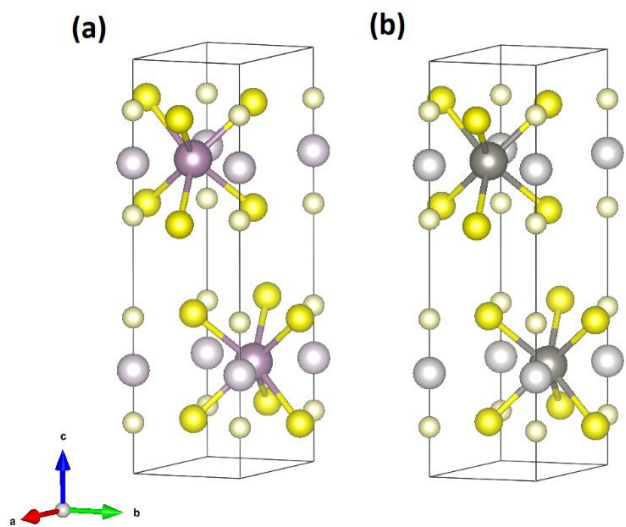

**Figure 2.** (a) Crystal structure of $MoS_2$ as determined by single-crystal X-ray diffraction. [25] Purple spheres represent Mo atoms, yellow spheres represent S. Lighter-shaded purple and yellow spheres indicate minorly-occupied Mo and S sites. (b) Crystal structure of $WS_2$ as determined by single-crystal X-ray diffraction. [26] Grey spheres represent W atoms. Lighter-shaded grey and yellow spheres indicate minorly-occupied W sites.

The SXRD patterns from three different planes (0kl, h0l and hk0) in the reciprocal lattice for both compounds are shown in **Figure 3**. Streaking, which would indicate the presence of a large fraction of stacking faults, is not clearly visible in these reciprocal lattice planes. This is consistent with the small fraction of stacking faults inferred from the structural refinements. **Figure 4** contains an illustration of one possible type of stacking fault that would lead to the observed partially-occupied sites in $MoS_2$; The stacking faults as envisioned can be interpreted as an occasional intergrowth of $3R-MoS_2$ in an otherwise 2H structure.

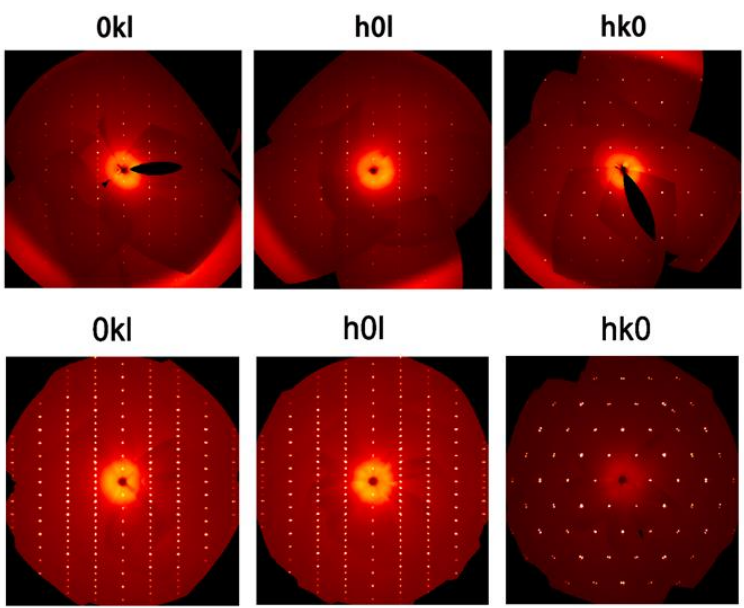

**Figure 3.** Top: Observed single crystal x-ray diffraction patterns for a crystal of $MoS_2$ in three different reciprocal lattice planes (*0kl*, *h0l* and *hk0*). Bottom: The equivalent patterns for $WS_2$.

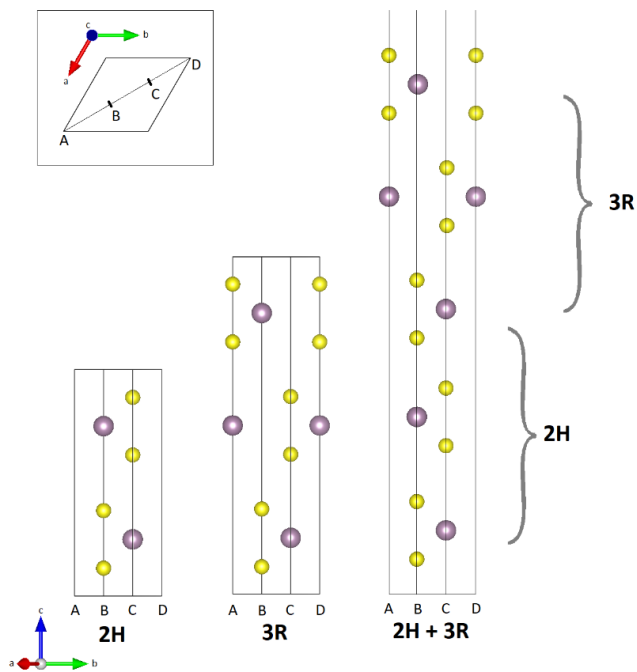

**Figure 4.** Illustration of one possible origin of the partially-occupied Mo and S sites in MoS$_2$. The structures of pristine 2H-MoS$_2$ [27] and 3R-MoS$_2$ [28] are shown, as well as a demonstration of how the partially-occupied sites can arise as a result of stacking faults in the form of a 3R intergrowth in the otherwise 2H structure. Inset: A diagram indicating the projection used.

Larger flakes of MoS$_2$ and WS$_2$ were inspected using SEM to determine whether they were true single crystals or merely well-defined agglomerations of smaller crystals. Images of two flakes of MoS$_2$ can be seen in **Figure 5**, and one flake of WS$_2$ in **Figure 6**. The lack of visible boundaries or seams in the surfaces of the flakes suggest that these samples are single crystals. Small hexagon-like growth features ("islands") can be observed on the surfaces of all MoS$_2$ flakes studied. A close-up image some of these hexagonal features can be seen in **Figures 5(c) and 7(b)**. **Figure 5(c)** is of particular interest, as the hexagonal island has itself developed a similar feature on its surface. The flakes of WS$_2$ do not appear to exhibit the same growth features. EDX measurements on several samples indicate that the flakes are pure MoS$_2$ and WS$_2$ respectively, and that no other elements are present in a measurable quantity.

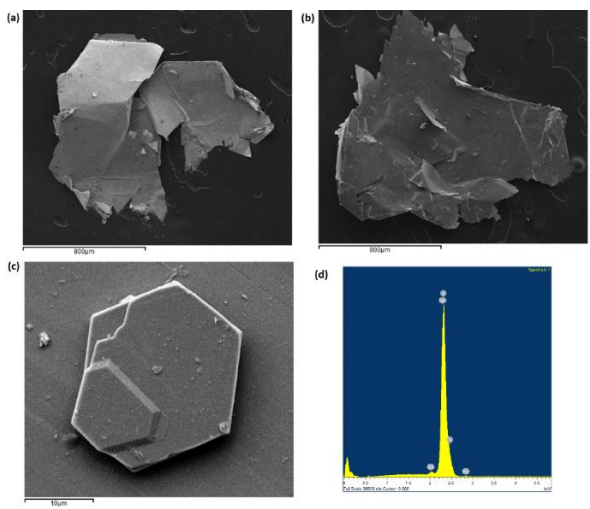

**Figure 5**. (a) and (b) SEM images of two flakes of MoS$_2$. (c) SEM image of hexagonal growth feature on surface of flake pictured in (b). (d) EDS spectrum on small region of feature pictured in (c).

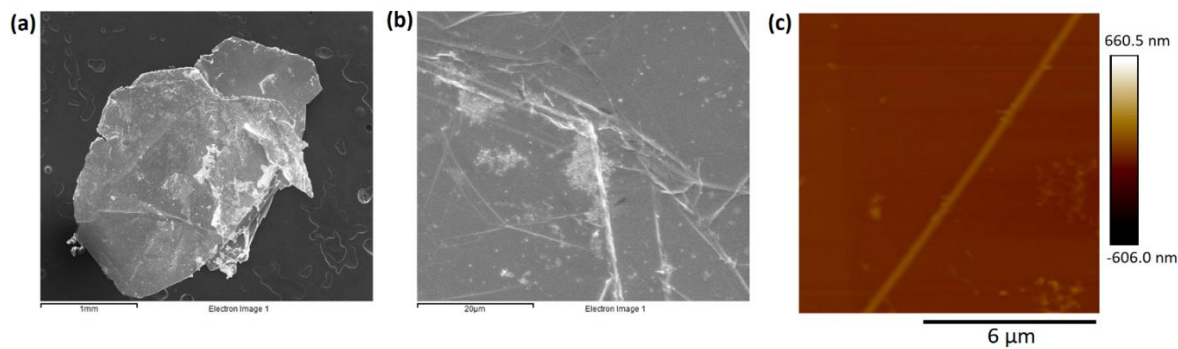

**Figure 6.** (a) SEM image of WS$_2$ flake. (b) Close-up of surface, showing hexagonal nature of layers but no small island growth features as in MoS$_2$. (c) AFM image of WS$_2$ surface, showing a layer edge but no clear indicators of crystal growth mechanism.

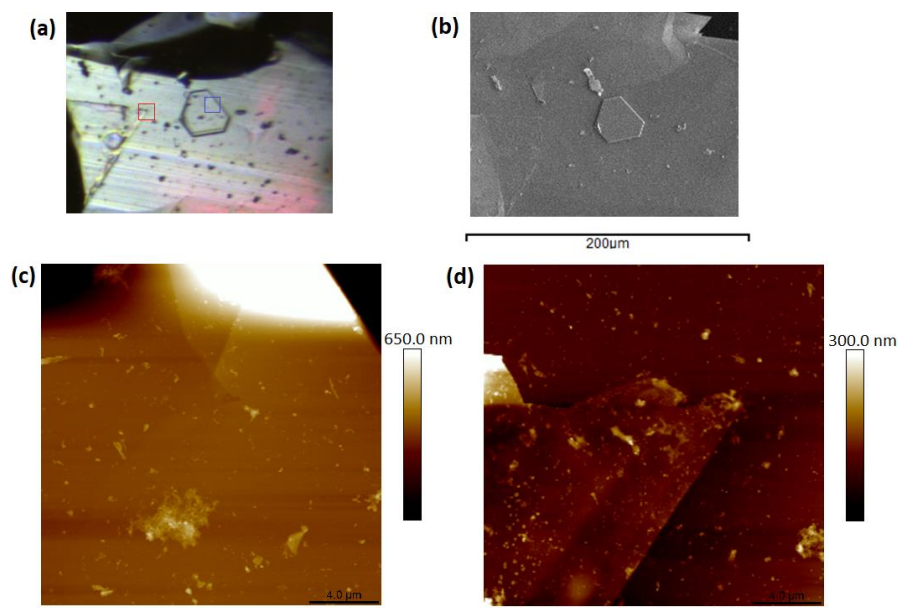

**Figure 7**. (a) Optical microscope image of the surface of a MoS$_2$ flake, showing visible striations across the surface, as well as a hexagonal surface feature. (b) SEM image of the same corresponding region. (c) AFM image of the surface of the hexagonal feature, indicated by the blue box in (a). (d) AFM image of the end of the triangular feature indicated by a red box in (a).

Results from the Sn-based flux growth of MoS$_2$ from Zhang et al. [16] indicated that their bulkier crystals result from screw-dislocation-driven (SDD) growth. In order to compare the growth mechanism of our samples, Atomic Force Microscopy (AFM) measurements were performed on several regions of a single-crystal flake of MoS$_2$. **Figure 7** depicts the results of AFM measurements on two regions of a crystalline flake; **Figure 7(c)** shows the surface of one of the hexagonal features, while **7(d)** depicts a striated triangular feature that was observed nearby. **Figures 7(a) and 7(b)** are optical and SEM images of the broader region, with specific areas of interest identified in **Figure 7(a)** by colored boxes. For clarity, **Figure 7(c)** has been cropped and rotated, but images have otherwise not been manipulated. While striations are visible on the crystal surface, there is no evidence of the spiral-like patterns typically associated with SDD growth. The presence of the hexagonal features on the crystal surface suggests instead that our crystals form via surface nucleation growth, which may explain the thinner nature of our crystals in relation to the bulkier samples of Zhang et al. A flake of $WS_2$ examined via AFM can be seen in **Figure 6(c)**. $WS_2$ flakes do not display any of the small hexagonal features, nor any of the spiral patterns that are indicative of SDD growth. The precise growth mechanism of these crystals therefore remains ambiguous.

In order to evaluate the intrinsic optical quality of the solution-transport-grown $MoS_2$, we prepared monolayers of $MoS_2$ by exfoliation and encapsulated between thin, insulating hexagonal boron nitride (hBN) on $SiO_2$ (285nm)/Si substrate as shown in **Figure 8(a)**. Photoluminescence (PL) spectra were taken at $T = 4$ K on the hBN/$MoS_2$/hBN heterostructure, as well as on similar heterostructures composed of different $MoS_2$ crystals available from commercial vendors (HQ Graphene and 2D Semiconductor). The PL spectra of these heterostructures, shown in **Figure 8(b)**, exhibit narrow neutral exciton ($X^0$) emission at energies between 1.93-1.95 eV due to a different dielectric environment, with a linewidth varying, depending on the sample used and the detection spot position. Notably, the exciton linewidth of our solution-transport-grown $MoS_2$ is around 4 meV, comparable to chloride-assisted vapor transport-grown $MoS_2$ [29] and two times sharper than that of either commercial sample (11.5 meV and 15.5 meV for HQ Graphene and 2D semiconductor respectively). This sharper excitonic emission in our monolayer $MoS_2$ suggests that our growth method can produce high optical quality without a pronounced degree of disorder that can lead to inhomogeneous broadening in the PL spectra.

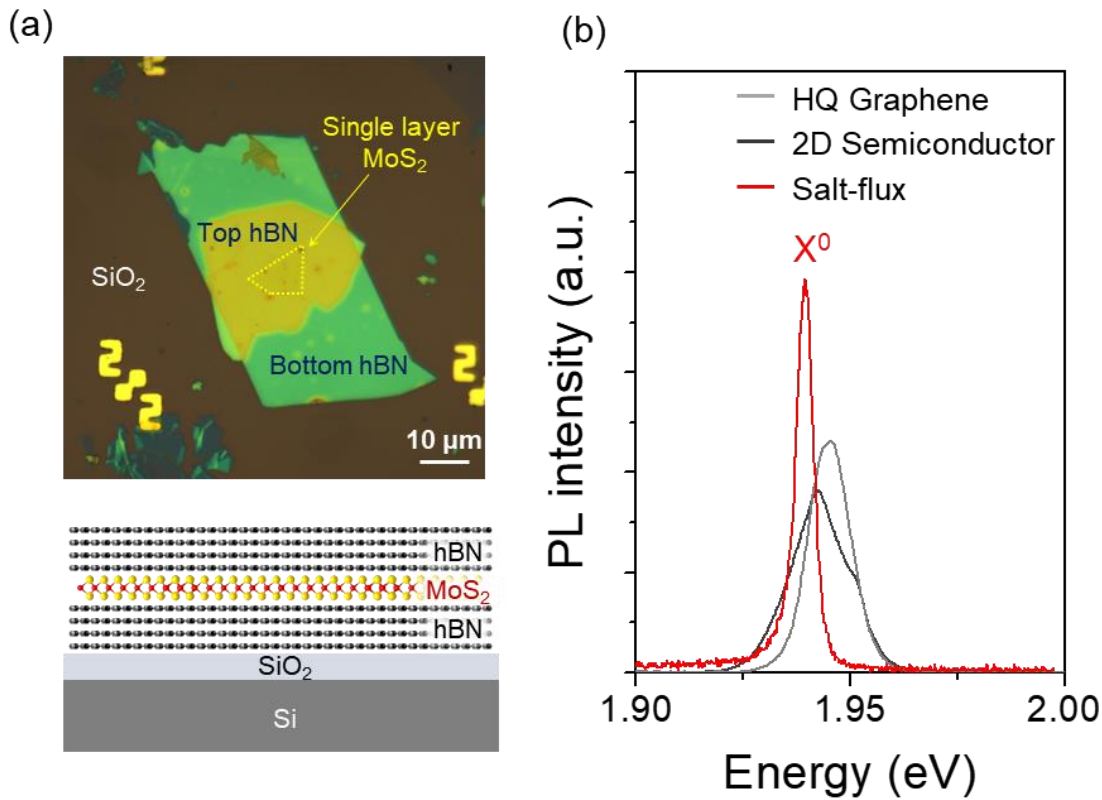

**Figure 8.** (a) (Top) Optical image and (Bottom) schematic of MoS$_2$ monolayer encapsulated between two hBN layers is placed on SiO$_2$ (285nm)/Si substrates. The dashed yellow lines show the outline of MoS$_2$. (b) Photoluminescence spectra at $T$ = 4 K for monolayer MoS$_2$ encapsulated with hexagonal boron nitride (hBN). The neutral exciton transition exhibits a linewidth of 4 meV for this solution-transport-grown monolayer MoS$_2$ (red) which is three times narrower compared to other commercial samples (gray and light gray).

**Conclusion**

Leaf-like single crystals of MoS$_2$ and WS$_2$ have been grown using a liquid-transport flux method, using a CsCl/NaCl eutectic mixture in a horizontal configuration. In contrast to a previously published tin-flux growth by Zhang et al. [16] the resulting crystals of MoS$_2$ are thin and have no clear signatures of screw-dislocation-driven crystal growth. The crystals are found to be higher-quality than those available commercially. We propose that this method may be preferable in instances where tin contamination is of significant concern, or screw dislocations are particularly undesirable. Crystals of $WS_2$ grown by the same mechanism do not show clear signs of surface nucleation or screw-dislocation-driven growth, leaving the precise growth mechanism ambiguous.

Aside from the sulfide compounds described here, early experiments suggest that this method may also be successfully applied to the growth of single crystals of the related transition metal diselenides $WSe_2$ and $MoSe_2$. Further study is currently underway to determine the general applicability of this technique to the broader family of TMDC compounds.

**Supporting information**

The following files are available free of charge.

CIF files for $MoS_2$ and $WS_2$.

Document containing 4 tables of crystallographic refinements and results for $MoS_2$ and $WS_2$.

**Accession Codes**

CCDC 1922046 and 1921982 contain the supplementary crystallographic data for $MoS_2$ and $WS_2$, respectively. This data can be obtained free of charge via [www.ccdc.cam.ac.uk/data_request/cif](www.ccdc.cam.ac.uk/data_request/cif) or by emailing [data_request@ccdc.cam.ac.uk](data_request@ccdc.cam.ac.uk), or by contacting The Cambridge Crystallographic Data Center, 12 Union Road, Cambridge CB2 1EZ, UK; fax: +441223336033.

**AUTHOR INFORMATION**

**Corresponding Author**

*F. Alex Cevallos, Department of Chemistry, Princeton University, Princeton, NJ, 08544. E-mail: [fac2@princeton.edu](fac2@princeton.edu)

**ORCID**

F. Alex Cevallos: 0000-0002-3459-0091

Shu Guo: 0000-0002-2098-8904

Giovanni Scuri: 0000-0003-1050-3114

**Author Contributions**

The manuscript was written through contributions of all authors. All authors have given approval to the final version of the manuscript.

**Acknowledgements**

This research was supported by the Gordon and Betty Moore Foundation EPiQS initiative, Grants GBMF-4412 and GBMF-4543. H.P. acknowledges support from the DoD Vannevar Bush Faculty Fellowship (N00014-16-1-2825).

**References and Footnotes**

# Single crystals of the layered dichalcogenides $MoS_2$ and $WS_2$ grown by liquid phase transport

*F. Alex Cevallos[1], Shu Guo[1], Hoseok Heo[2], Giovanni Scuri[2], You Zhou[2], Jiho Sung[2], Takashi Taniguchi[3], Kenji Watanabe[3], Philip Kim[4], Hongkun Park[2], and Robert J. Cava[1]*

[1] Department of Chemistry, Princeton University, Princeton NJ 08542

[2]Department of Chemistry and Chemical Biology, Harvard University, Cambridge MA 02138

[3]National Institute for Materials Science, 1-1 Namiki, Tsukuba 305-0044, Japan

[4]Department of Physics, Harvard University, Cambridge MA 02138

**Table S1.** Crystal data and ambient temperature crystal structure refinements for $MoS_2$.

| Formula | $MoS_2$ |
|---|---|
| *formula mass(amu)* | 160.06 |
| *crystal system* | hexagonal |
| *space group* | $P6_3/mmc$ |
| *a*(Å) | 3.1601(10) |
| *c*(Å) | 12.288(4) |
| *V*(Å$^3$) | 106.27(6) |
| *Z* | 2 |
| *T*(K) | 300(1) |
| *ρ(calcd)(g/cm$^3$)* | 5.002 |
| *λ* (Å) | 0.71073 |
| *F(000)* | 148 |
| *θ(deg)* | 3.32- 33.20 |
| *Index ranges* | $-4 \leq h \leq 4$ |
| | $-4 \leq k \leq 4$ |
| | $-18 \leq l \leq 18$ |
| Cryst size (mm$^3$) | $0.051 \times 0.041 \times 0.02$ |
| *μ*(mm$^{-1}$) | 7.648 |
| *Final R indices ($R_1/\omega R_2$)* | 0.0144/0.0367 |
| *R indices (all data) ($R_1/\omega R_2$)* | 0.0166/0.0369 |
| *Residual electron density/ (eÅ$^{-3}$)* | (-0.802) - 0.873 |

**Table S2**. Wyckoff positions, coordinates, occupancies, and equivalent isotropic displacement parameters for $MoS_2$.

| Atoms | Wyck. Site | x/a | y/b | z/c | S.O.F. | $U_{eq}$ |
|---|---|---|---|---|---|---|
| Mo1 | 2c | 1/3 | 2/3 | 1/4 | 0.9686(14) | 0.00425(15) |
| Mo2 | 2b | 0 | 0 | 1/4 | 0.0314(14) | 0.00425(15) |
| S1 | 4f | 1/3 | 2/3 | 0.62278(8) | 0.9686(14) | 0.00620(19) |
| S2 | 4e | 0 | 0 | 0.377(2) | 0.0314(14) | 0.00620(19) |

**Table S3.** Crystal data and ambient temperature crystal structure refinements for $WS_2$.

| Formula | $WS_2$ |
|---|---|
| *formula mass(amu)* | 247.98 |
| *crystal system* | hexagonal |
| *space group* | $P6_3/mmc$ |
| *a*(Å) | 3.1599(4) |
| *c*(Å) | 12.3554(17) |
| *V*(Å$^3$) | 106.84(2) |
| *Z* | 2 |
| *T*(K) | 300(1) |
| *ρ(calcd)(g/cm$^3$)* | 7.708 |
| *λ* (Å) | 0.71073 |
| *F(000)* | 212 |
| *θ(deg)* | 6.61-31.96 |
| *Index ranges* | $-4 \leq h \leq 4$ |
| | $-4 \leq k \leq 4$ |
| | $-18 \leq l \leq 18$ |
| Cryst size (mm$^3$) | 0.04 × 0.04 × 0.02 |
| *μ*(mm$^{-1}$) | 55.530 |
| *Final R indices ($R_1/\omega R_2$)* | 0.0119/0.0247 |
| *R indices (all data) ($R_1/\omega R_2$)* | 0.0143/0.0255 |
| *Residual electron density/ (eÅ$^{-3}$)* | (-1.709) - 1.170 |

**Table S4**. Wyckoff positions, coordinates, occupancies, and equivalent isotropic displacement parameters respectively for $WS_2$.

| Atoms | Wyck. Site | x/a | y/b | z/c | S.O.F. | $U_{eq}$ |
|---|---|---|---|---|---|---|
| W1 | 2c | 2/3 | 1/3 | 3/4 | 0.9921(12) | 0.00616(16) |
| W2 | 2b | 0 | 0 | 3/4 | 0.0079(12) | 0.00616(16) |
| S1 | 4f | 1/3 | 2/3 | 0.62277(13) | 0.992 | 0.0077(3) |
| S2 | 4e | 0 | 0 | 0.380(13) | 0.0079 | 0.0077(3) |